\documentclass{esannV2}
\usepackage[dvips]{graphicx}
\usepackage[latin1]{inputenc}
\usepackage{amssymb,amsmath,array}
\usepackage{acronym}
\usepackage{color}
\usepackage{bm}
\usepackage{subcaption}
\usepackage{multirow}
\usepackage{url}

\acrodef{PDE}[PDE]{Partial Differential Equation}
\acrodefplural{PDE}[PDE]{Partial Differential Equations}

\acrodef{RBC}{Rayleigh-B\'enard Convection}
\acrodef{DNS}{Direct Numerical Simulation}
\acrodefplural{DNS}[DNS]{Direct Numerical Simulations}
\acrodef{ROM}[ROM]{Reduced Order Model}
\acrodefplural{ROM}[ROM]{Reduced Order Models}
\acrodef{FNO}{Fourier Neural Operator}
\acrodef{DMD}{Dynamic Mode Decomposition}
\acrodef{LRAN}{Linearly-Recurrent Autoencoder Network}

\acrodef{API}[API]{Application Programming Interface}
\acrodefplural{API}[APIs]{Application Programming Interfaces} % Define plural form

\begin{document}

%style file for ESANN manuscripts
\title{Solving Turbulent Rayleigh-B\'enard Convection using Fourier Neural Operators}

%***********************************************************************
% AUTHORS INFORMATION AREA
%***********************************************************************
\author{Michiel Straat$^1$, Thorben Markmann$^1$ and Barbara Hammer$^1$
%
% Optional short acknowledgment: remove next line if non-needed
\thanks{The authors acknowledge financial support by the project "SAIL: SustAInable Life-cycle 
of Intelligent Socio-Technical Systems" (Grant ID NW21-059A), which is funded by the 
program "Netzwerke 2021" of the Ministry of Culture and Science of the State of 
North Rhine Westphalia, Germany.}
%
% DO NOT MODIFY THE FOLLOWING '\vspace' ARGUMENT
\vspace{.3cm}\\
%
% Addresses and institutions (remove "1- " in case of a single institution)
1- Bielefeld University - Center For Cognitive Interaction Technology \\
Inspiration 1, 33619 Bielefeld - Germany
}
%***********************************************************************
% END OF AUTHORS INFORMATION AREA
%***********************************************************************

\maketitle

\begin{abstract}
We train \ac{FNO} surrogate models for \ac{RBC}, a model for convection processes that occur in nature and industrial settings. We compare the prediction accuracy and model properties of \ac{FNO} surrogates to two popular surrogates used in fluid dynamics: \ac{DMD} and the \ac{LRAN}. We regard \acp{DNS} of the \ac{RBC} equations as the ground truth on which the models are trained and evaluated in different settings. The \ac{FNO} performs favorably when compared to the \ac{DMD} and \ac{LRAN} and its predictions are fast and highly accurate for this task. Additionally, we show its zero-shot super-resolution ability for the convection dynamics. The \ac{FNO} model has a high potential to be used in downstream tasks such as flow control in \ac{RBC}.
\end{abstract}

\section{Introduction}
%Artificial Intelligence techniques are increasingly used to complement or replace traditional numerical methods for solving Partial Differential Equations (PDE).
The fluid dynamics field benefits greatly from the application of AI techniques in several respects \cite{vinuesa22,markmann24}: Accelerating \ac{DNS}, improving model accuracy, and developing \acp{ROM}, which are models that decompose the dynamics in its most prominent features, akin to PCA and non-linear autoencoders. In this work, we focus on surrogate models, which are models that replace the \ac{DNS} for predicting future roll-outs of the system. They can be used in a purely data-driven manner on observation data without requiring equations or model parameters. Surrogate models are significantly faster than \ac{DNS}, even when taking training data generation and model training into account \cite{li21}. Hence, they are highly suitable for parameter studies in dynamical systems and solving inverse problems. Additionally, their differentiability makes them applicable to implementing control schemes.
% Other examples are AI models for complex phenomena such as turbulence \cite{duraisamy_2019}.
%These models may also be used in the planning phases of model-based RL methods for fluid flow control.

However, surrogate models are often dependent on the resolution of the training data. To address this, the \ac{FNO} model was recently introduced \cite{li21} as a versatile network architecture that directly learns the complex-valued coefficients of convolutional Fourier-space filters. The main benefit of operator models is their function space representation \cite{straat_2020}, which makes them suitable for learning solution operators of \acp{PDE} that generalize to different spatial resolutions. The \ac{FNO} performed particularly well for learning solution operators of \acp{PDE}, as was demonstrated on fluid flow described by Navier-Stokes in a chaotic regime \cite{li21}.
%A similar approach of learning on intrinsically functional data in a prototype-based framework was presented in \cite{straat_2020} for classification problems.

In this work, we employ the \ac{FNO} model for the first time to study its effectiveness as a surrogate for turbulent convection dynamics. Specifically, we address its effectiveness as a function of the amount of turbulent flow in the system. Convection is described by the \ac{RBC} \ac{PDE}, which models a fluid in a box that is heated from below, which causes the fluid to rise to the top of the box. Increasing the heat at the bottom makes these upward flows more turbulent. Convection is a widespread phenomenon in nature (atmosphere, Earth's mantle, oceans) and industrial settings (e.g. silicon waver production), and studying it using novel AI methods has the potential to improve applications in meteorology and the chemical industry.
We aim at approximating a solution operator (or surrogate) for \ac{RBC} up to highly turbulent regimes using the \ac{FNO} and compare its performance to two models that we chose due to their popularity in fluid dynamics: \ac{DMD} and \ac{LRAN} \cite{markmann24,williams_2016}. These methods aim to find a linear dynamical system in a latent space of system measurements that approximate the non-linear dynamics.

%TODO This could be moved to a discussion at the end, because it contains some specific things about RBC and it sort of fits content-wise also at the end.
In the literature, modeling of the convective field in \ac{RBC} was done in \cite{pandey22} using an autoencoder and a GRU in the latent space. Here, we focus on resolution-independent operator models for the entire state of \ac{RBC} that include the fluid velocities and the temperature fields. In \cite{kontolati2024}, the \ac{FNO} and a DeepONet operator model in the latent space were compared in predicting the initial motion for one time unit in the \ac{RBC} system starting from a no-motion initial condition. In contrast, we assume that convective cells have already formed and we model their dynamic patterns for long time windows and varying degrees of turbulence.

%The paper is structured as follows: first, we will briefly discuss the \ac{DNS} used for obtaining the ground truth dynamics and introduce the architecture with our choices of hyperparameters in Sec.~\ref{sec:methodology}. Then we will show and discuss the results in Sec.~\ref{sec:results} and Sec.~\ref{sec:discussion}, respectively. Points for future work are raised in \ref{sec:conclusion}.

\section{Methodology} \label{sec:methodology}
\subsection{Data generation using simulations}
All the surrogate models used in this paper are trained fully data-driven on observations. Usually, those observations are taken from sensors in experiments \cite{howle1997}, atmosphere, or industry. In this work, we rely on computer simulations of convection in 2D to generate the training data.
We used a numerical solver for the \ac{RBC} equations from the Shenfun \cite{shenfun} package on a 2D rectangular spatial domain\footnote{Spatial dimensions: Horizontal $x \in [0, 2\pi]$, vertical $y \in [-1, 1]$, discretized to $96\times 64$ uniform grid points. Bottom temperature boundary condition (BC) $T_H=2$ and top BC $T_C=1$. Zero velocity BC at top and bottom, periodic BCs at left and right. Solver timestep $dt=0.1$. Our code repository provides further details and equations:  \url{https://github.com/SAIL-project/RBC-FNO-Surrogate}.}.
The simulation yields images of the state over time and each "pixel" location has the local fluid's temperature value and velocity vector.
%Introducing the time variable $t$, the solution to the PDE consists of the velocity and temperature field that are denoted with $\bm{u}(x,y,t)$ and $T(x,y,t)$, respectively.
%We put a temperature of $T_H=2$ at the lower boundary and $T_C=1$ at the upper boundary ($\Delta T= T_H - T_C = 1)$. Then, the boundary- and initial conditions were:
% \[
% \begin{cases}
%     \bm{u}(x,-1,t) = \bm{u}(x,1,t) = \bm{0}, & \text{i.e. no-slip horizontal BC} \\
%     T(x,-1,t) = T_H, \, T(x, 1, t) = T_C, \\
%     \bm{u}(0,y,t) = \bm{u}(2\pi, y, t), \, T(0,y,t)=T(2\pi,y,t), & \text{i.e. periodic vertical BC} \\
%     \bm{u}(x,y,0) = \bm{0}, \\ 
%     T(x,y,0) = T_H - \frac{y+1}{2} \Delta T + 0.001 \nu(x,y),
% \end{cases}\,
% \]
%where $\nu(x,y)$ is a standard Gaussian random field.

The \textit{Rayleigh number} $Ra$ is a key system parameter that determines the amount of convective turbulence in the system.
%is the ratio of buoyancy forces (due to the temperature gradient $\Delta T$) to the stabilizing effects of the fluid's viscosity and thermal diffusivity. Informally, $Ra$ is a measure of the vigor of convective turbulence \cite{pandey22}.
For our study,
%we fixed $Pr=0.7$ and
we varied $Ra \in \{1e5, \, 1e6, \, 2e6, \, 5e6\}$, which resulted in four settings starting from moderate to high turbulence.
The simulations were run for 200 initial time units to let the system transition from the no-motion state to convection and subsequently, 250 time units were recorded for the training data.
In total, we generated 25 episodes for each Rayleigh number starting from slightly different random initial conditions. We divided the episodes into $[15, 5, 5]$ episodes for training, validation and testing, respectively. For the quantitative evaluation of the predictions, we calculated a Normalized Root Sum of Squared Errors (NRSSE)\footnote{For the evaluation of all methods, we selected ten random starting points in each test episode for which we recurrently applied the models that performed best in the validation to predict a further time window of length 30, after which we calculated the average NRSSE with respect to the ground truth over all selected test windows.
}:
%i.e., the norm of the difference between the predicted state $\hat{\bm{x}}$ and the \ac{DNS} ground truth state $\bm{x}$, normalized by the norm of the ground truth, i.e.:
\begin{equation} \label{eq:NRSSE}
    NRSSE = ||\hat{\bm{x}} - \bm{x}|| \, / \, ||\bm{x}||\,,
\end{equation}
where $\hat{\bm{x}}$ is the predicted state and $\bm{x}$ is the ground truth state.

\begin{figure}
    \centering
    \begin{subfigure}[b]{0.5\textwidth}
        \includegraphics[width=\linewidth]{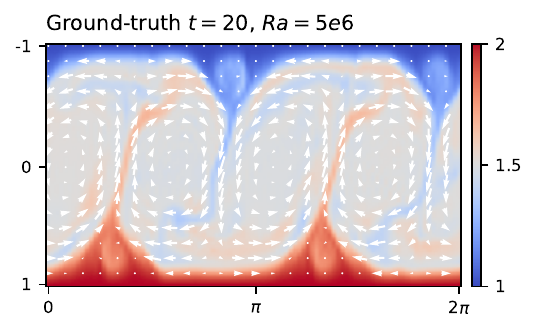}
    \end{subfigure}%
    \begin{subfigure}[b]{0.5\textwidth}
        \includegraphics[width=\linewidth]{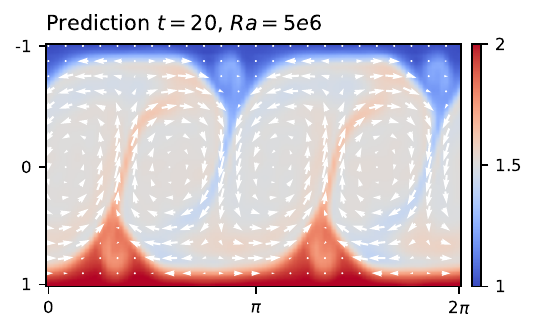}
    \end{subfigure}
    \caption{\textit{Left}: A ground truth at $t=20$ from a random test starting point that we label $t=0$. \textit{Right}: the field as predicted by FNO-3D. Color: temperature field, arrows: velocity field.}
\end{figure}

\subsection{Fourier Neural Operator}
Operator methods use trainable functional representations to learn a solution operator of a dynamical system in a fully data-driven way and invariant to resolution.
The \ac{FNO} is a specific operator architecture proposed in \cite{li21}, in which the training of global convolutional filters takes place in Fourier space, where convolutions are implemented as multiplication. Similar to general neural networks, the linear processing in a layer is followed by a non-linearity, and several of these Fourier layers are used sequentially.
%\textit{Fourier layers} that each performs a global convolution implemented as a product in Fourier space and a local linear operation, followed by a non-linear activation function.
The multiplication in Fourier space truncates the data to a predefined number of lower-frequency Fourier modes. An MLP at the start of the network lifts the state to a higher-dimensional number of hidden channels, and an MLP at the end projects back to the target dimension. See Fig.~2 in \cite{li21} for an illustrative depiction of the architecture.

We applied the FNO-3D variant that learns an operator that maps 3D functions to 3D functions by learning a series of spatiotemporal filters.
%In training, we experimented with mapping 3D volumes of time window  to volumes of the same shape, leaving a time of $0.5$ between individual snapshots and a time of $1$ between individual input-output pairs.
%For instance, in the case of $T=10$, this means we train the operator on pairs with a resolution of $64 \times 96 \times 20$ (product of 2D spatial and time resolution).
We performed searches over data and architecture parameters\footnote{We experimented with mapping volumes of time window $T \in \{10, 15, 20\}$ as input and output with $\Delta t = 0.5$ between individual snapshots. We performed a parameter search over the number of Fourier layers $\{8, 16, 32\}$, lower Fourier modes in the signal $\{8, 16, 32\}$, hidden channels $\{16,32,64\}$ and hidden units in the lifting and projection MLP $\{16,32,64\}$.}. Subsequently, we extracted 3300 input-output pairs from the 15 training episodes, 1100 pairs from the five validation episodes, and 1100 pairs from the five testing episodes.
%We identified approximately optimal sets of hyperparameters by a Bayesian search in the following space: the number of Fourier layers $f_1 \in \{4, 6, 8\}$, the number of lower Fourier modes to retain $f_2 \in \{8, 16, 32\}$, the number of hidden channels on which the Fourier layers work $f_3 \in \{16,32,64\}$ and the number of hidden units in both the lifting- and projection MLPs $f_4 \in \{16,32,64\}$, the learning rate in ADAM from the range $lr \in [10^{-4}, 10^{-2}]$ and the values of the aforementioned parameter $T$.

\subsection{Koopman architectures}
Koopman methods are popular in fluid dynamics and aim to extract a linear dynamical system in a theoretically infinite-dimensional space of \textit{observables}, which are non-linear measurements of the system. Due to linearity, it is possible in some cases to decompose the dynamics into Koopman eigenvalues and Koopman modes (which act like eigenvectors).
Next, we briefly discuss two methods that are used in fluid dynamics to extract linear dynamical systems in practice.
\subsubsection{Dynamic Mode Decomposition}
\ac{DMD} provides a linear system on a finite-dimensional subspace,
%As the number of possible observables is infinite, the Koopman operator is infinite-dimensional.
which can grow exponentially with the dimension of the state space and the complexity of the dynamics.
For this reason, the kernel \ac{DMD} leverages the kernel trick to implicitly compute inner products in the high-dimensional observable space \cite{williams_2016}.
The SVD is used to fit\footnote{We performed a parameter search over the width of the Gaussian kernel $\sigma \in [9.0, 100.0]$ and the length of the fitting window $T \in [20, 100]$.} the DMD on a window and the extracted Koopman modes and eigenvalues are used to predict the future evolution. Similar to ARMA models, refitting is necessary whenever making predictions.

\subsubsection{Linearly Recurrent Autoencoder Network}
The \ac{LRAN} differs from the \ac{DMD} in that it uses an autoencoder network architecture to learn 1.) the observable functions and 2.) the linear transition matrix in observable space, both simultaneously by end-to-end training using gradient descent\footnote{We performed a parameter search over the latent dimension of the autoencoder $[10, 1000]$ and the length of the windows during training $[5, 30]$. The autoencoder was pre-trained on single snapshots of the system to accelerate the training of the whole architecture later.}. Contrary to the FNO which takes a spatiotemporal volume as input, the \ac{LRAN} takes a single input snapshot and autoregressively predicts arbitrarily-sized prediction windows.
See Fig.~2 in \cite{markmann24} for an illustrative depiction of the architecture and \cite{otto2019} for an in-depth treatment of the \ac{LRAN}.

% parameters are the size of the Koopman subspace, i.e. the latent dimension of the autoencoder, and the window length of state sequences
% the lran only takes a single state and predicts arbitrarily many following states autoregressively
% we perform a hyperparameter search for Koopman subspace sizes k1={} and the length of state windows during training k2={} similar to [quote markmann]
% autoencoder is pretrained on single state reconstructions to accelerate the training of the LRAN

\section{Results and Discussion} \label{sec:results}
\begin{figure}
\centering
\includegraphics[width=\linewidth]{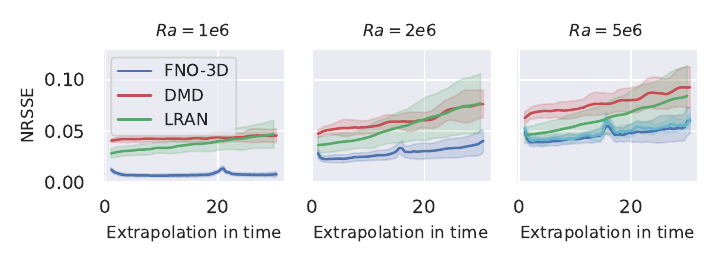}
\caption{Evaluation of the three models as an average error \eqref{eq:NRSSE} computed over 50 random starting points (10 random points in each of the 5 test episodes) for increasing Rayleigh number (see figure titles). The Cyan line for $Ra=5e6$ shows the same FNO model but evaluated on data with double the spatial resolution.} \label{fig:results_predictions}
\end{figure}

\begin{table}[htb]
    \centering
    \footnotesize
    \begin{tabular}{|m{0.08\linewidth}||m{0.08\textwidth}||m{0.12\textwidth}|m{0.12\textwidth}|m{0.1\textwidth}|m{0.1\textwidth}|m{0.1\textwidth}|}
        \hline
        Method & Overall NRSSE & Training Time [hrs] & Prediction Time per window [s] & Memory [MB] & Resolution Invariant & Eigen Analysis \\
        \hline
        FNO  & \textbf{0.021} & $\sim$1-2    & \textbf{0.45}      & 3037              & \textbf{yes}      & no            \\
        LRAN & 0.042 & $\sim$1-2    & 0.47               & \textbf{1080}              & no                & \textbf{yes} \\
        DMD  & 0.054 & -            & 3.16               & 1675                 & yes*              & \textbf{yes} \\
        \hline               
        DNS  & - & -             & 50                 & 23      & yes*              & -            \\
        \hline     
    \end{tabular}
    \caption{Model Comparison regarding prediction accuracy, computational complexity and their properties. *Note that DMD has to be refitted on each window, i.e., DMD does not generalize to other initial conditions or system parameters.}
    \label{tab:comparison_props_models}
\end{table}

% \begin{table}[htb]
%     \centering
%     \footnotesize
%     \begin{tabular}{|c|c|c|c|c|c|c||c|c||c|c|c|c|c|c|}
%         \hline
%         $Ra$ & $f_1$ & $f_2$ & $f_3$ & $f_4$ & $lr$ & $T$ & $d_1$ & $d_2$ & $k_1$ & $k_2$ & $lr$ \\
%         \hline
%         $1e5$ & 6 & 16 & 32 & 16 & 6e-4 & 10 & $10$ & 20 & 25 & 15 & 5e-4 \\
%         $1e6$ & 6 & 32 & 16 & 32 & 4e-4 & 20 & $82$ & 90 & 50 & 15 & 5e-4  \\
%         $2e6$ & 6 & 16 & 16 & 16 & 9e-4 & 15 & $85$ & 30 & 50 & 15 & 5e-4  \\
%         $5e6$ & 6 & 16 & 16 & 32 & 3e-4 & 15 & $51$ & 100 & 50 & 15 & 5e-4  \\
%         \hline
%     \end{tabular}
%     \caption{Results from separate parameter searches optimizing for test set performance for the three methods FNO-% 3D ($f_i$), DMD ($d_i$) and LRAN ($k_i$). The $f_i$, $d_i$ and $k_i$ are defined in Sec.~\ref{sec:methodology}.}
%     \label{tab:configurations_FNO}
% \end{table}

The main result is given in Fig.~\ref{fig:results_predictions}, which shows the performance on the 30-second window prediction task averaged over randomly chosen starting points from the test episodes. We observed, also by qualitative inspection, that all methods performed very well on the moderately turbulent cases up to $Ra=10^6$, with the FNO-3D scoring best. For the higher turbulent cases, the error increased over the predicted window. Here, too, the FNO-3D was the most accurate, although the Koopman methods were competitive.

Table~\ref{tab:comparison_props_models} lists the overall accuracy and other properties of the models.
The FNO and \ac{LRAN} converged in a few hours on an Nvidia A40 and generalized well to the test episodes with only the cost of forward passes, making these methods suitable for planning in model-based RL methods and parameter studies in dynamical systems.
%Predicting a single test window on the CPU for $Ra=2*10^6$ took $0.45$, $0.47$, $3.16$, and $50$ seconds for FNO-3D, LRAN, DMD, and the DNS (time step: $dt=0.025$), respectively.
Although the FNO is 111 times faster than the DNS, its memory consumption can be substantial, as displayed by a high memory consumption compared to the other methods. A benefit of the FNO is that the trained model generalizes to other resolutions in the working phase. As we show for $Ra=5e6$ in Fig.~\ref{fig:results_predictions}, it performs well on higher resolution data than it was trained on, without additional computational cost, while the other methods need retraining on expensive \ac{DNS} simulations for higher resolutions.
%In contrast, a slower method such as \ac{DMD} only stores eigenvectors and eigenvalues.

The \ac{DMD} is appropriate for modeling periodic patterns, which are indeed present in \ac{RBC}. Although this explains the good accuracy, it should be noted that a re-fitting is necessary each time when making predictions. The fitting has a complexity of $O(n*m^2)$, where $n$ is the number of dimensions and $m$ is the number of snapshots in the window ($n=3*64*96,m \leq 200$). This is likely too slow in situations where a large number of forward evaluations is necessary. %which may not be suitable in cases where many evaluations are needed.

\section{Conclusion} \label{sec:conclusion}
In this work, we investigated FNO-3D as a surrogate model for different levels of convective turbulence and compared the results with Koopman methods that are popular in the fluid dynamics field. All studied surrogate models were fully data-driven. At all levels of turbulence, the FNO-3D was superior concerning accuracy and zero-shot generalization to higher resolutions. The current FNO models can be used as a fast surrogate for the DNS, and can also be used in similar situations when only measurement data is available. Our future work will incorporate the fast FNO-3D predictions in a model-based RL framework for flow control in convection systems. We also aim to study more realistic 3D settings of convective flows.

These fast and accurate AI-based solutions for convection dynamics have a large potential to improve tasks in weather modeling and the chemical industry.

% ****************************************************************************
% BIBLIOGRAPHY AREA
% ****************************************************************************

\begin{footnotesize}

\bibliographystyle{unsrt}
\bibliography{bibliography}

\end{footnotesize}

% ****************************************************************************
% END OF BIBLIOGRAPHY AREA
% ****************************************************************************

\end{document}